\documentclass[12pt]{article}

\usepackage{scicite}

\usepackage{times}

\usepackage[dvipsnames]{xcolor}%added by Ziqiang
\definecolor{burntorange}{rgb}{0.8392 0.2745 0.0353} %added by Ziqiang
\usepackage{siunitx}
\usepackage{graphicx}
\usepackage{amssymb}
\usepackage{fdsymbol} %%%This looks important for the figure symbols
\usepackage{amsmath}
\newenvironment{sciabstract}{%
\begin{quote} \bf}
{\end{quote}}

\title{Direct imaging of polymer filaments pulled from rebounding drops}

\author
{Zi Qiang Yang,$^{1}$ Peng Zhang,$^{2}$ Meng Shi,$^{1}$ Ali Al Julaih$^{1}$, \\
Himanshu Mishra,$^{2}$ Enzo Di Fabrizio,$^{3}$ Sigurdur T. Thoroddsen$^{1\ast}$\\
\\
\normalsize{$^{1}$Division of Physical Sciences and Engineering, King Abdullah University of Science}\\
\normalsize{ and Technology (KAUST), Thuwal, 23955-6900, Saudi Arabia}\\
\normalsize{$^{2}$Division of Biological and Environmental Science and Engineering, King Abdullah}\\
\normalsize{University of Science and Technology (KAUST), Thuwal, 23955-6900, Saudi Arabia}\\
\normalsize{$^{3}$Dept. of Applied Science and Technology, Politecnico di Torino, Torino, 10129, Italy}\\
\\
\normalsize{$^\ast$To whom correspondence should be addressed; E-mail:  sigurdur.thoroddsen@kaust.edu.sa}
}

\date{}

\begin{document} 


\baselineskip24pt
\maketitle 

\begin{sciabstract}
  Polymer filaments form the foundation of biology from cell scaffolding to DNA. 
Their study and fabrication play an important role in a wide range of processes from tissue engineering to molecular machines.
We present a simple method to deposit stretched polymer fibers between micro-pillars.  
This occurs when a polymeric drop impacts on and rebounds from an inclined superhydrophobic substrate.  
It wets the top of the pillars and pulls out liquid filaments which are stretched and can attach to adjacent pillars leaving minuscule threads, with the solvent evaporating to leave the exposed polymers.  
We use high-speed video at the microscale to characterize the most robust filament-forming configurations, by varying the impact velocity, substrate structure and inclination angle, as well as the PEO-polymer concentration. 
Impacts onto plant leaves or randomized nano-structured surface leads to the formation of a branched structure, through filament mergers at the free surface of the drop.  
SEM shows the deposition of filament bundles which are thinner than those formed by evaporation or rolling drops. 
Raman spectroscopy identifies mode B stretched DNA filaments from aqueous-solution droplets. %single-strand molecules from acquous DNA-solution droplets.
\end{sciabstract}

% In setting up this template for *Science* papers, we've used both
% the \section* command and the \paragraph* command for topical
% divisions.  Which you use will of course depend on the type of paper
% you're writing.  Review Articles tend to have displayed headings, for
% which \section* is more appropriate; Research Articles, when they have
% formal topical divisions at all, tend to signal them with bold text
% that runs into the paragraph, for which \paragraph* is the right
% choice.  Either way, use the asterisk (*) modifier, as shown, to

\section*{Introduction}

Methods to produce and collect polymer filaments are of primary importance in biology, tissue engineering, medicine, pharmacology and textiles \cite{goyanes2014, van2000DNA, zhang2014textiles, soto2018, stassi2019}.
% \cite{soto2018, van2000DNA, goedert2017, goyanes2014, brown2016melt, zhang2014textiles, stassi2019}.
% [1] S. Stassi, M. Marini, M. Allione, S. Lopatin, D. Marson, E. Laurini, S. Pricl, C.F. Pirri, C. Ricciardi, E. Di Fabrizio, Nat Commun, 10 (2019) 1690.
% [2] P. Zhang, M. Moretti, M. Allione, Y. Tian, J. Ordonez-Loza, D. Altamura, C. Giannini, B. Torre, G. Das, E. Li, S.T. Thoroddsen, S.M. Sarathy, I. Autiero, A. Giugni, F. Gentile, N. Malara, M. Marini, E. Di Fabrizio, Commun Biol, 3 (2020) 457.
% [3] Van Brabant, A. J., Stan, R. & Ellis, N. A. Dna helicases, genomic instability, and human genetic disease. Annual review of genomics and human genetics 1 (2000).
% [4] Zhang, D. Advances in lament yarn spinning of textiles and polymers (Elsevier, 2014).
% \cite{chiti2017, soto2018, goedert2017, van2000DNA}
% Gerhold, D., Rushmore, T. & Caskey, C. T. (1999) Trends Biochem. Sci. 24 , 168–173. pmid:10322428
% Cuzin, M. (2001) Transfus. Clin. Biol. 8 , 291–296. pmid:11499980
% Michalet, X., Ekong, R., Fougerousse, F., Rousseaux, S., Schurra, C., Hornigold, N., van Slegtenhorst, M., Wolfe, J., Povey, S., Beckmann, J. S., et al. (1997) Science 277 , 1518–1523. pmid:9278517
% van Brabant, A. J., Stan, R. and Ellis, N. A.
% (2000), ‘DNA helicases, genomic instability,
% and human genetic disease’, Ann. Rev.
% Genomics Hum. Genet., Vol. 1, pp. 409–459.
Several attempts have been reported to obtain single molecular-level polymer filaments like DNA suspended between micro-pillars, using methods like stamp peeling \cite{lebofsky2003, guan2005}, evaporation of fakir drops \cite{wang1998, quere2002fakir, gentile2012, marini2014raman, ciasca2014, marini2015} and drop sliding \cite{su2012small, su2012elaborate, ciasca2013}.
% F. De Angelis, F. Gentile, F. Mecarini, G. Das, M. Moretti, P. Candeloro and E. Di Fabrizio, “Breaking the diffusion limit with super-hydrophobic delivery of molecules to plasmonicnanofocusing SERS structures”, Nature Photonics 5(11), 682-687 (2011).

\begin{figure}[t]
\includegraphics[width=0.7\textwidth]{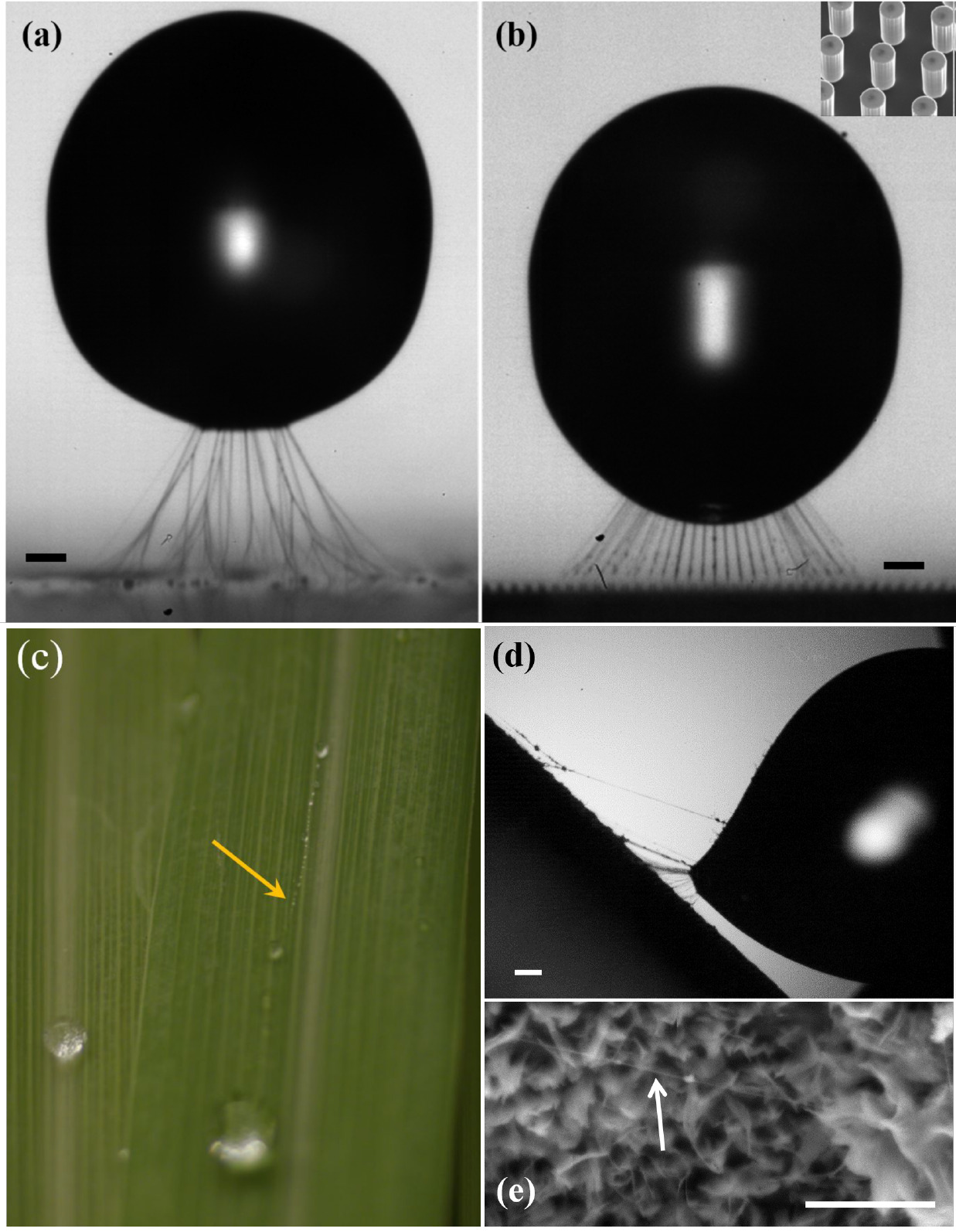}
\caption{Polymer filaments pulled from rebounding drops after impact on a nano-particle-coated surface in {\bf (a)} and on micro-pillars in {\bf (b)}.
In (a) the glass substrate has four layers of ``glaco'' nanoparticle coating \cite{vakarelski2012}, with impact conditions $D=1.40$ mm, $U=0.36$ m/s, $Re=463$ \& $We=2.64$,
while in (b) the micro-pillars are coated with FDTS, for $D=1.42$ mm, $U=0.42$ m/s; $Re=552$ \& $We=3.67$.  The inset shows an SEM image of the pillars, which are 50 $\mu$m tall and spaced by 50 $\mu$m.
For the random-roughness glaco-surface the filaments become branched, while for the pillars regularly-spaced linear filaments are pulled out.
\textcolor{black}{The scale bars are 200 $\mu$m long.}% with 20 $\mu$m diameter.} 
For more details, see Supplementary Videos 1 \& 2.
{\bf (c)} Similar polymer drops leave polymer filaments (arrow) when bouncing from superhydrophobic leaves of lemon-grass.
(d) Video frame of the configuration in (c).  {\bf (e)} SEM image of filaments left behind on the leave.  Scale bar is 5 $\mu$m.}
\label{Fig_1}
\end{figure}

\textcolor{black}{Herein, we demonstrate a robust approach to produce and collect highly ordered arrays of polymer nanostrands with well-defined length and orientation.
This involves impacting a drop containing polymers on a microfabricated superhydrophobic pillared surface,
where the rebouncing pulls out the filaments and then deposits them onto the pillars.
We show that both PEO and DNA filaments can become thinner by drop impact, than by the evaporative technique.
We use time-resolved 100,000 fps high-speed video imaging, 
on the micro-scale, to study the details of this process. 
Different filament structures can be pulled from the drop during the rebouncing, depending on whether the substrate has random roughness or regular pillars, as shown in Fig. \ref{Fig_1}. 
This raises the prospect of targeted pesticide delivery, especially in dry climates, when molecular structures on these filaments could embed targeted molecules, protecting against the arrival of pests.  
The details of the filament pulling could also help design optimal spraying techniques to increase droplet retention on superhydrophobic leaves.  Our preliminary experiments with lemongrass, show filament formation for repeated bounces of a micro-drop from the leaf surface, as shown in Fig. \ref{Fig_1}(c).}
%Both PEO and DNA filaments are tested to demonstrate the feasibility of this method.

Droplets can bounce at low impact velocities due to the presence of an unbroken lubricating air film between the impinging droplet and the surface, as has been studied extensively using interferometry \cite{Ruiter2015b, kolinski2014drops, van2012direct, van2014microstructures, quere2002fakir, LiThoroddsen2015}.
Superhydrophobic surfaces with specific surface structures like various pillar arrangements also enable drop bouncing \cite{barthlott1997, richard2000bouncing,reyssat2006bouncing,bartolo2006bouncing, deng2009, kwon2013, kwon2011} with reduced contact time \cite{bird2013reducing,liu2014pancake}. 
Chen et al. \cite{chen2018} and Li et al. \cite{li2020promoting} have previously observed these filaments and found them to delay the drop retraction.
%used a high-speed camera to study the impact of aqueous polymer droplets on solid superhydrophobic surface, and they found that viscoelastic filaments are generated to delay drop retraction. 
However, their 20 $\mu$m/px resolution could not capture the details of the finest filaments observed herein.
%%%%%%%%%%%%%%%%%%%%%%%%%%%%%%%%%%%%%%%%%%%%%%%%%%%%%%%%%%%%%%%%%%%%%%%%%%%%%%
%  FIGURE 2
\begin{figure*}
\includegraphics[width=0.693\textwidth]{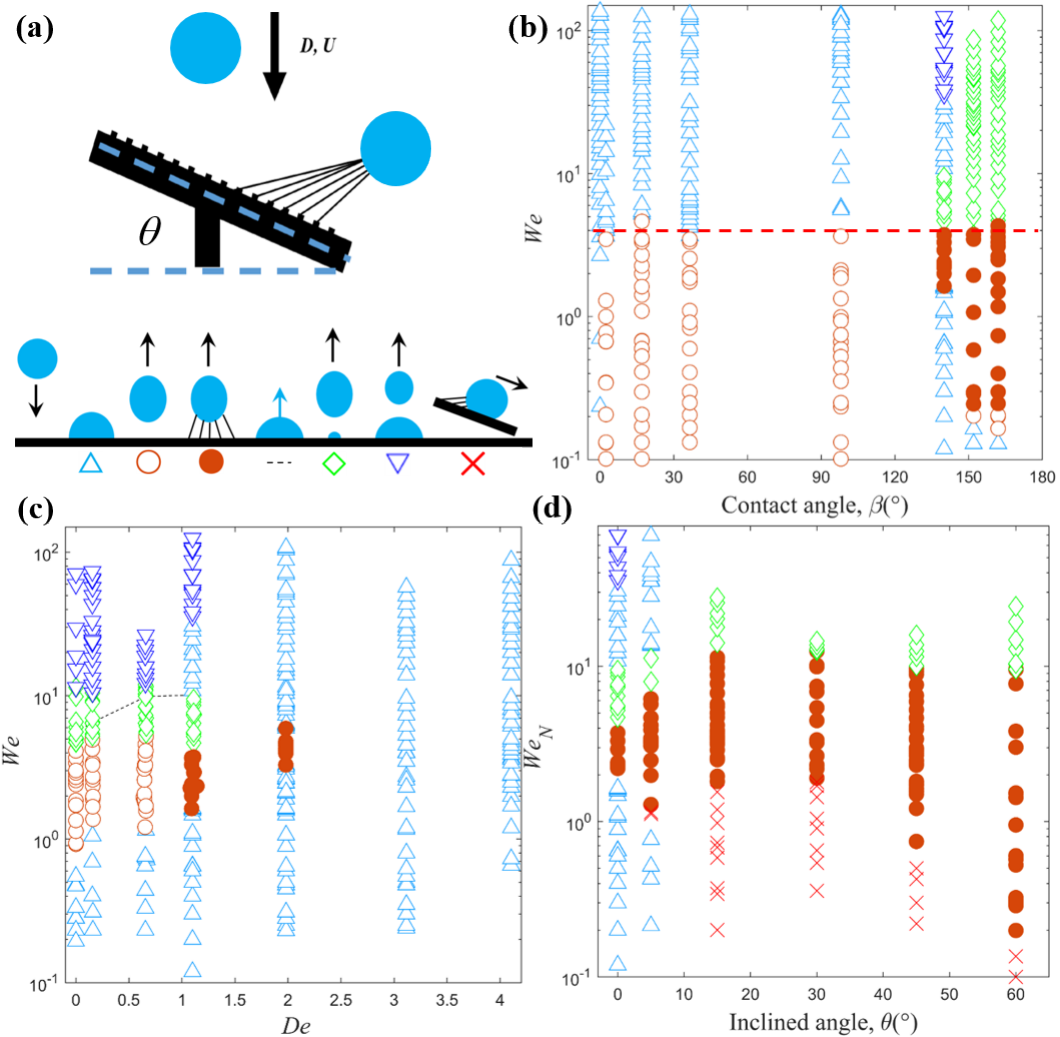}
  \caption{{\bf Parameter regimes for the impact of a polymeric water drop:} bouncing with filaments (\textcolor{burntorange}{$\lgblkcircle$}) and bouncing without filament formation (\textcolor{burntorange}{$\bigcirc$}). 
   The other symbols correspond to different impact outcomes:
  %(\textcolor{red}{$\bigcirc$}) droplet rebounding; 
  (\textcolor{cyan}{$\triangle$}) droplet deposition on the solid surface; 
  (\textcolor{green}{$\Diamond$}) partial rebounding with liquid blobs left on the pillars;
  (\textcolor{black}{$--$}) singular jet from the drop apex;
  (\textcolor{blue}{$\triangledown$}) droplet deposition and then separated due to vibration; 
  (\textcolor{red}{$\times$}) droplet slides along the surface, leaving filaments.
  The range of impact velocities is presented in terms of the Weber number, $We=\rho D U^2/\sigma$.
  {\bf (a)} Sketch of drop impact onto inclined micro-pillared substrate.
  {\bf (b)}  The influence of surface micro-structure and contact angle of the various surfaces, during normal impacts of 100 ppm PEO droplets, with $D\simeq 1.5 \pm 0.1$ mm on a horizontal substrate.  Each data column corresponds to different surface treatment or micro-structure with the corresponding values of the contact angle $\beta$:
  Cylindrical superhydrophilic micro-pillars: $\beta = 0^o$; 
  Molecularly smooth mica sheet: $\beta = 2.4^o$; 
  Smooth silica treated with plasma: $\beta = 17.4^o$; 
  Flat silica: $\beta = 36.5^o$; 
  Silica coated with FDTS: $\beta = 98^o$; 
  Cylindrical micro-pillars coated with FDTS: $\beta = 140^o$; 
  One time Glaco-coated \textcolor{black}{glass} surface: $\beta = 152^o$; 
  Four times Glaco-coated \textcolor{black}{glass} surface: $\beta = 162^o$. 
  {\bf (c)}  Effect of PEO concentration for a drop impacting on a horizontal surface with cylindrical micro-pillars coated with FDTS.  The PEO concentration is characterized by the Deborah number $De$, going from low to high at 0, 10, 50, 100, 200, 400, 1000 ppm, i.e. from left to right in the figure.
  {\bf (d)} Influence of the surface inclination angle $\theta$.  Here we use the Weber number based on the normal component of the impact velocity, $U_N=U\, cos\, \theta$.}
%   \textcolor{violet}{the red droplets seem deformed, please correct}
\label{Fig_05}
\end{figure*}

%  V^* = ( 0.073 * 50x10^-6 / ( 1000 x (50x10^-6)^2 ) )^0.5 = 
% (0.073/50x10^-3)^0.5 = ( 0.073 / 0.050 )^0.5 = 1.2 m/s 
% We = 1000x(1.4*10^-3)x1.2^2/0.073 = 28
%%%%%%%%%%%%%%%%%%%%%%%%%%%%%%%%%%%%%%%%%%%%%%%%%%%%%%%%%%%%%%%%%%%%%%%%%%%%%%%%%%%%

%{\bf Results:}
We first map the bouncing regimes of a polymer drop impacting on different solid substrates, while keeping the PEO concentration fixed at 100 ppm.
%Drop bouncing series are observed for 100 ppm PEO droplets impacting on different substrates. 
The surfaces are characterized by the effective contact angle $\beta$, listed in the caption of Fig. \ref{Fig_05}.  They
include flat mica sheet, flat silica, silica surface treated with plasma, silica surface coated with FDTS, one \& four times Glaco-nanoparticle-coated surfaces and surface with cylindrical micro-pillar array (height $h=50\; \mu$m; diameter $d=20\; \mu$m; spacing $w=50\; \mu$m) coated with FDTS, see also Supplementary Table 1.

The phase-diagram in Fig. \ref{Fig_05}(b) sums up the bouncing behavior of 100 ppm PEO droplets, of diameter $D$, impacting on these surfaces at different velocities $U$, characterized in terms of the Weber number $We=\rho D U^2/\sigma$, where $\rho$ is the liquid density and $\sigma$ the surface tension. 
Starting with the uncoated micropillar surface ($\beta \simeq 0^o$), \textcolor{black}{a complete deposition is observed over the whole range of $We$, because the liquid wets the whole superhydrophilic substrate immediately following contact with the pillars}.  
% This is caused by the surface being superhydrophilic at 0$^{\circ}$ contact angle, which makes the liquid wet the surface immediately, not allowing the lubricating air layer, or surface tension to pull the free surface back upwards from the pillars. 
On the other hand, for all of the smooth flat surfaces ($\beta = 2^o \rightarrow 100^o$), %mica, silica treated with plasma, silica, silica coated with FDTS surfaces, 
the drop bounces within a very similar range starting at $We_{min} \sim$ 0.1 to $We_{max} \sim 4$ (dashed red line in Fig. \ref{Fig_05}(b)),
above this value the drop deposits on the surface. %3.4 ... 4.6. 
This is true even for the hydrophilic substrate and agrees well with earlier results \cite{Ruiter2015b,kolinski2014drops},
% Kolinski JM, Mahadevan L, Rubinstein SM. 2014a. Drops can bounce from perfectly hydrophilic surfaces. Eur. Phys. Lett. 108:24001
% The maximum impact speed for to bounce is 0.48 m/s for PEO droplets at 100 ppm concentration, corresponding to $We_{max}$ = 4,
indicating the 100-ppm PEO drops are not in direct contact with the solid, with a stable lubricating air-layer between drop and substrate.
This lack of contact prevents any filament formation.
For the rough or structured superhydrophobic surfaces ($\beta \ge 140^o$) the outcomes are more varied. 
The most prominent change, is the repeatable formation of polymer filaments during the rebounding.  Furthermore, for the lowest impact velocities, i.e. the smallest $We$, direct deposition occurs.  Above this we observe repeatable
rebounding which always generates fiber filaments. For the FDTS-coated micropillars ($\beta=140^o$) this rebounding window is narrowest between $We= 2.19 - 3.74$, followed by partial rebounding and deposition at larger $We$.

It is therefore clear that in rebounding cases, there must be penetration through the air-layer and local wetting to pull liquid filaments from the drop.  Whether this liquid filament forms into a polymer filament is shown below to depend on the polymer concentration.
%\textcolor{green}{This is most likely due to the largest protrusions of this surface. The FDTS coated micropillars always demonstrate filament generation in all rebounding and partial rebounding cases, which is most likely also can relate to the regularity of the structure.} 

%%%%%%%%%%%%%%%%%%%%%%%%%%%%%%%%%%%%%%%%%%%%%%%%%%%%%%%%%%%%%%%%%%%%%%%%%%%%%%
%  FIGURE 3
\begin{figure*}
\begin{center}
(a) \includegraphics[width=0.95\textwidth]{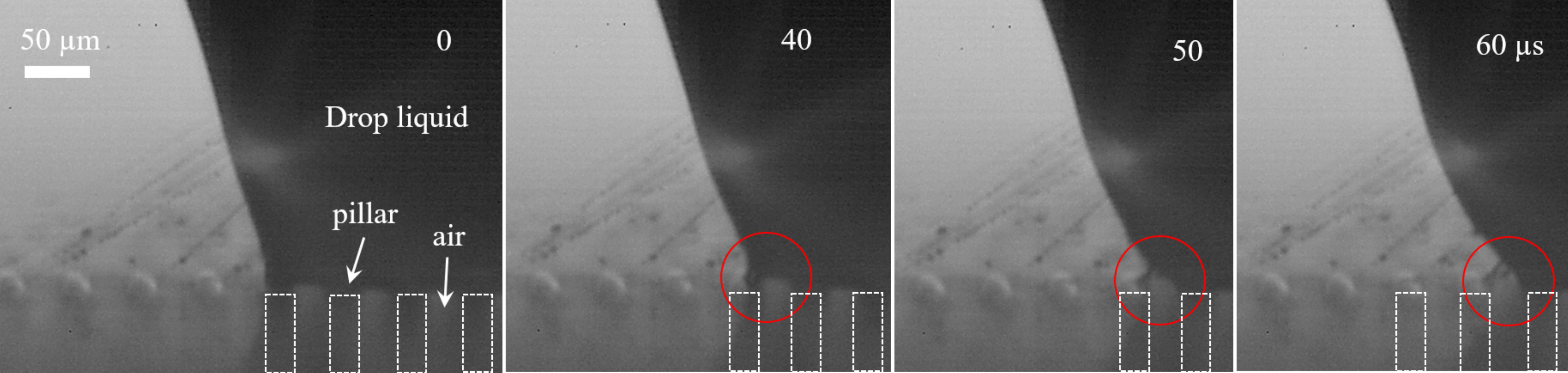}\vspace{0.1in}\\
(b) \includegraphics[width=0.95\textwidth]{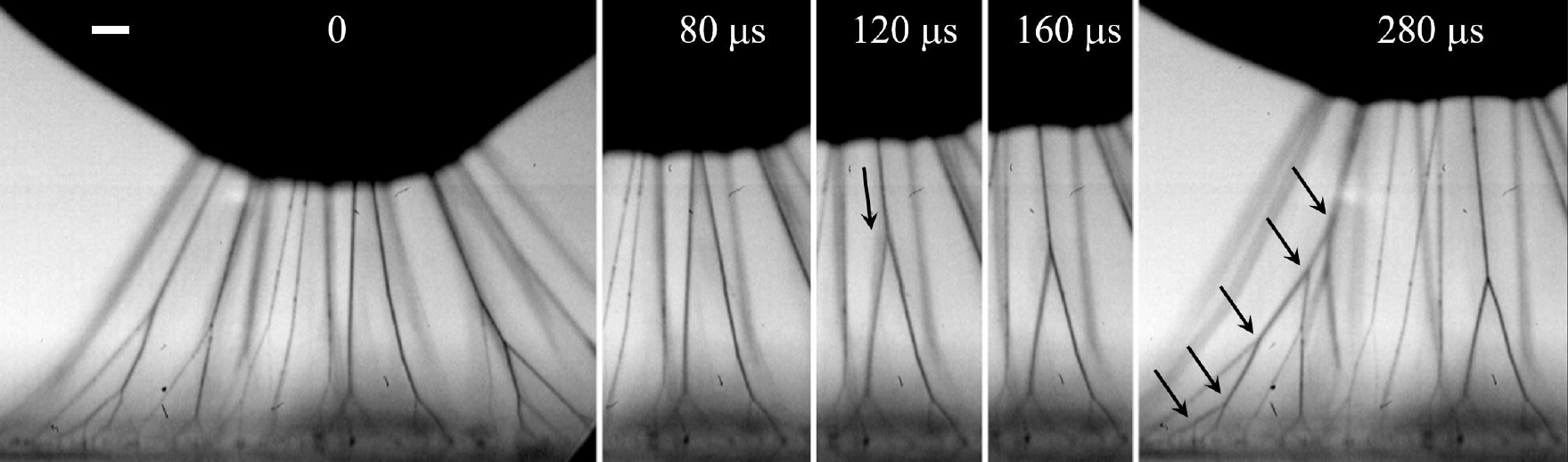}\vspace{0.1in}\\
\end{center}
\caption{Filament formation for impact on horizontal substrates.  (a)  The process of liquid pinning on top of the pillars and pulling out of filaments, as the rim of liquid body moves to the next pillars during the retraction,
marked by the red circles.  The impact conditions are the same as in Fig. \ref{Fig_1}(b).
%\textcolor{red}{For $D=??$ mm, $U= ??$ m/s, $We=??$.  Same conditions as in Fig. 1(b) ?}
(b)  Branching of filaments during rebound from a random nano-particle coated glass surface, under the same impact conditions as in Fig. \ref{Fig_1}(a). 
% and a concave liquid surface was formed due to the retracted pulling of the main liquid body; III, capillary bridge is stretched and the PEO filaments are generated between outer pillars and main droplet liquid body. 
% Schematic figure for this process is drawn in figure 3 (b);
%IV, the filaments are stretched longer between outer pillars and droplets and the contact line moves to outer pillars to repeat this process;
% (b) The sketch of the generation of the filaments responding to step III in (a).
%(c) The tracking of location and moving velocity of the contact line between the $\mu$-pillars and droplet.
% The images are acquired with an ultra-fast camera (Phantom V2511) at 100,000 fps by Leica lens (2.0 × 9.2 × 1.6×).
The first frames are chosen as time reference and the scale bars are 50 $\mu$m long.
See also Supplementary movies 3 \& 4.  The first arrow indicates the merging motion of the two filaments.
The arrows in the last panel identify the nodal points of the multiply-branched structure.
}
\label{Fig_03}
\end{figure*}

Similar progression of outcomes is observed for the random roughness of the nano-particle-coated Glaco surfaces. Both one and four coatings ($\beta = 152^o$ \& $162^o$) show direct deposition for the lowest impact velocities.  
Furthermore, we find a narrow range of bouncing without filaments, which is followed, starting at $We \simeq 0.2$, by a large range of rebounding with polymer filaments. This filament-formation regime covers a much wider range of $We$ than for the regular pillars.  Again, for $We > 4$ one sees partial rebounding, with prominent liquid blob left behind (see inset sketch in Fig. \ref{Fig_05}(a)).
We note that the critical $We=4$ needed for the bottom of the drop to penetrate the air film is not changed for textured or flat surfaces, as marked by the dashed red line in Fig. \ref{Fig_05}(b).
Finally, for even larger $We$, we see partial rebounding over a large range of $We$.  Further increase in $We$ above the region shown in Fig. \ref{Fig_05}(b) splashing begins from the rapidly expanding lamellae \cite{Josserand2016}.

These results show that the most regular filaments occur for impacts on the superhydrophobic micropillar array, which we now focus on.
Figure \ref{Fig_05}(c) shows how changing the concentration of PEO affects the impact dynamics on these pillars. 
We characterize this concentration in terms of the Deborah number, which is the ratio of time-scales $De = \tau_{E}/\tau_{R}$, where $\tau_{E}$ is the extensional relaxation time of the polymers, while Rayleigh capillary-inertial timescale $\tau_{R}=\sqrt{\rho \, (D/2)^{3}/\sigma}$ marks the hydrodynamic reaction time from surface tension.
Starting with pure water we see a large rebounding region while deposition occurs for both lower and higher impact velocities.
For low PEO concentrations of 10 \& 50 ppm, while similar to water, the rebounding regime reduces, replaced by partial rebounding for $We > 5$ and then above $We \simeq 10$ larger wetting and deposition occurs with vertically ejected droplets. 
However, filaments are not observed during these rebounds,
perhaps due to insufficient polymer concentrations to stabilize the liquid filaments emanating from the contacts with the pillar surfaces.
%The partial rebounding continues to grow with 50 ppm concentration, and this can be linked to the fact that the PEO concentration is not enough to form filaments which act as pulling forces to the surface thus minimizing the flying time and height of the drop. 
%This explanation can be observed when looking at the 100 ppm PEO concentration where the filaments start to generate and it is noticeable that the partial rebounding window gets much smaller. The 100 ppm demonstrates a good regime since it consists of reasonable rebounding height and amount of filament generation. Considering the importance of the rebounding height, 200 ppm demonstrates a remarkable filament generation compared to 100 ppm, with no observed partial rebounding, however, a low rebounding height with tiny pinch-off droplets are observed which both lead to irregular filament formation, and difficulty in the filament collection scenario. 
Drops of both 100 and 200 ppm concentrations show clear $We-$ranges of rebounds with filaments.  For 200 ppm the region moves to higher $We$, thereby requiring larger impact velocities.  The pulling of more filaments reduces the rebounding height (see Suppl. Fig. {\bf S3}) and partial rebound no longer occurs.   The 200 ppm also generates thicker filaments with more droplets along them, which leads to a more irregular filament formation. Considering the importance of the rebounding height, for filament-collection scenarios, we therefore conclude that 100 ppm is the optimal concentration for our purposes. 
For even higher concentrations of PEO, of 400 ppm and 1000 ppm in \textcolor{black}{Fig. \ref{Fig_05}(c)}, the rebounding regime disappears and deposition rules the entire range of $We$. 

In the parameter space of Figure panels \textcolor{black}{\ref{Fig_05}(b,c)} the impacts were all perpendicular to the horizontal substrates.  To deposit and study the filaments the drop cannot be allowed to return straight back down, so we instead incline the substrate and let the drop bounce downwards to leave the filaments attached to the pillars, as sketched in \textcolor{black}{Fig. \ref{Fig_05}(a)}.
Figure \ref{Fig_05}(d) shows the results for the 100 ppm PEO drop impacting on the pillars, over a large range of inclination angles.
Here we use the normal component of the impact velocity, $U_N = U\, cos\, \theta$, in the Weber number $We_N$.  The inclined substrates all show larger regions of filament formation than for first impact on horizontal pillars, $\theta = 0$.  For low impact velocities the drops tend to slide along the pillars \textcolor{black}{before rebounding}.  
For $\theta > 10^o$ this sliding leaves filaments and occurs over a large range of low impact velocities, reminiscent of the pure sliding tested by \cite{su2012small, su2012elaborate, ciasca2013}.  At slightly higher $U$, for $We_N \sim 1$, the rebounding with filaments starts and persists until $We_N \simeq 12$, above which partial rebound occurs.  While similar dynamics occur for $\theta$ between $15^o-60^o$, for consistency we limit the detailed study below primarily to the $45^o$ angle.

\begin{figure*}
\begin{center}
\includegraphics[width=0.8\textwidth]{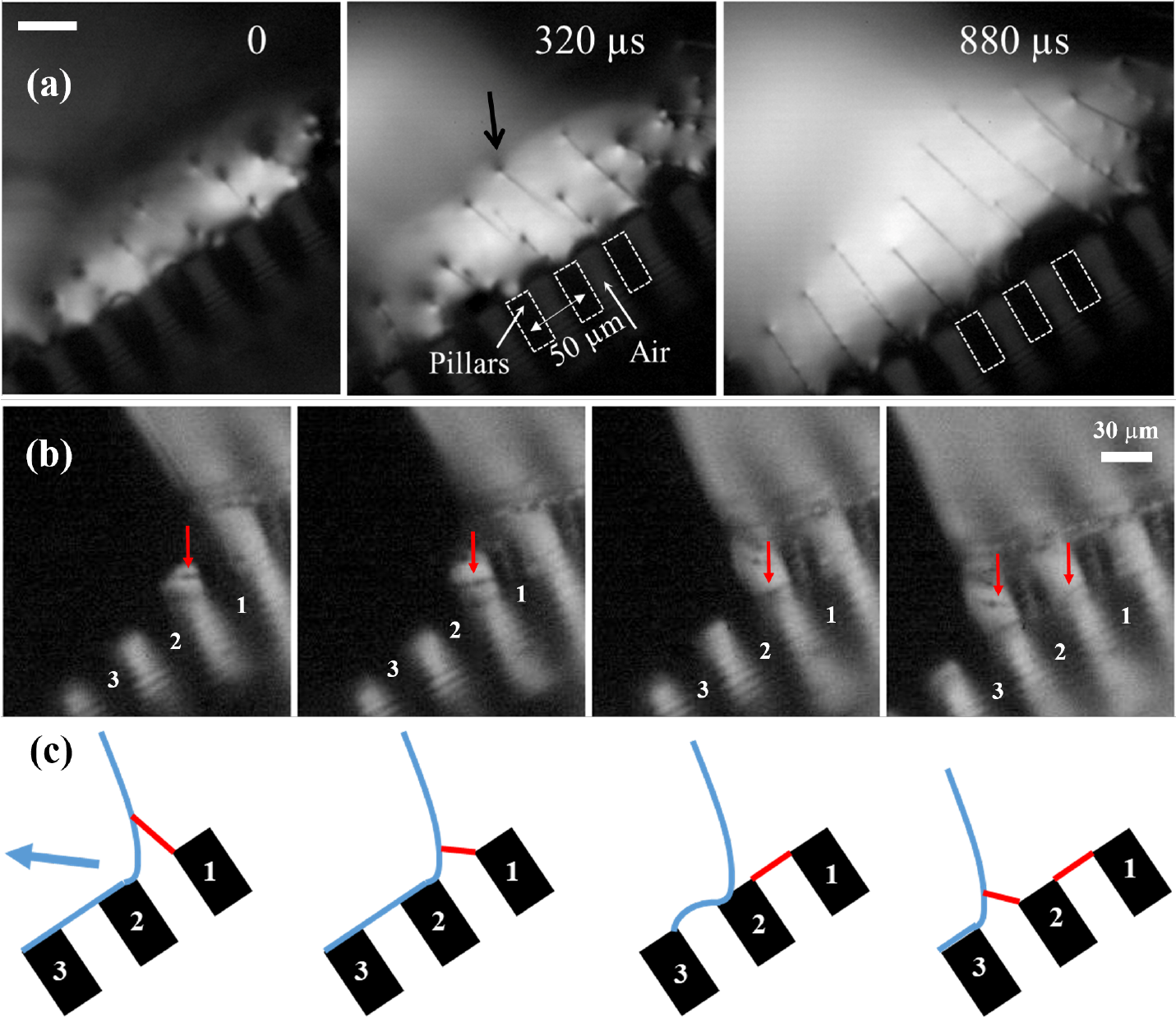}
\end{center}
\caption{Close-up video frame of filaments generation attached on the top of two pillars from impinging droplets of aqueous PEO solution on hydrophobic surfaces for 100 ppm at angle 30$^{\circ}$.
The impact condition is $D=1.62$ mm, $U=0.38$ m/s, $Fr= 9$, $We= 3$.
A small conical region where the filament connects to the free surface is marked by the black arrow.
The red arrows point out the generated filaments attached on the pillars.
The blue arrow note the drop bouncing direction.
The scale bars are 30 $\mu$m long.
Frames from a video clip taken at 25 kf.p.s.
See also Supplementary movies 5 and 6.}
\label{Fig_04}
\end{figure*}

We now study the details of the filament formation using time-resolved high-speed video at micron-level spatial resolution, from different viewing angles (see Supplemental).  Figure \ref{Fig_1} showed overall snapshots after the drops have risen far above the surface, with long filaments reaching from drop to substrate.  Figures \ref{Fig_03} and \ref{Fig_04} show close-up video frames of this process for horizontal and inclined surfaces respectively.  During the retraction of the edge of the drop leading to the rebound, in Fig. \ref{Fig_03}(a), we see the free surface of the drop attached to the top of a pillar, being pulled away leaving the filament.  
%%%%%%%%%%%%%%%%%%%%%%%%%
We also analyzed the location and moving velocity of the contact line between the micro-pillars and droplet in \textcolor{black}{Fig. \ref{Fig_03}(a,b)}, which shows a clear stepwise evolution of the pinning dynamics and the process of filaments generation and stretch, corresponding to the images in Fig. \ref{Fig_03}(a). The average retraction velocity is around 0.2 m/s while the velocity jumps to a higher value of 1.24 m/s when the droplet liquid jumps from an outer pillars onto an inner one.
The cylindrical pillars are marked by dashed white boxes and air is trapped between the pillars and the droplet. 
%%%%%%%%%%%%%%%%%%%%%%%%%%%%%%%%%%%%%%%%%%%%%%%%%%%%%%%%%%%%%%%%%%%%%
The diameter of this liquid bridge reduces from about 50 $\mu$m to $\sim 5 \; \mu$m in only 60 $\mu$s, suggesting an effective stretching rate of \textcolor{black}{$10^6$ s$^{-1}$}, thereby straightening the polymers.

%%%%%%%%%%%%%%%%%%%%%%%%%%%%%%%%%%%%%%%%%%%%%%%%%%%%%%%%%%%%%%%%%%%%%%%%%%%%%%
%  FIGURE 4

Figure \ref{Fig_04}(a) shows a close-up sideview of filaments being pulled out of the drop's free surface, during rebounding from a \textcolor{black}{$30^o$} inclined substrate.  The arrow in the second panel points out where a filament attaches to the liquid surface.  One notices a small conical region around the filament %\textcolor{red}{marked by the black arrow}, 
where the two surface curvatures adjust.  This matching region appears to reduce in size with time, as the filament thins.
The size of the cones is much smaller than the spacing of the filaments,
which corresponds directly to the distance between pillars, just like observed for the horizontal impacts in Fig. \ref{Fig_1}(b).

%%%%%%%%%%%%%%%%%%%%%%%%%%%%%%%%%%%%%%%%%%%%%%%%%%%%%%%%%%%%%%%%%%%%%%%%%%%%%%
%  FIGURE 5
\begin{figure*}[t]
	\includegraphics[width=0.98\textwidth]{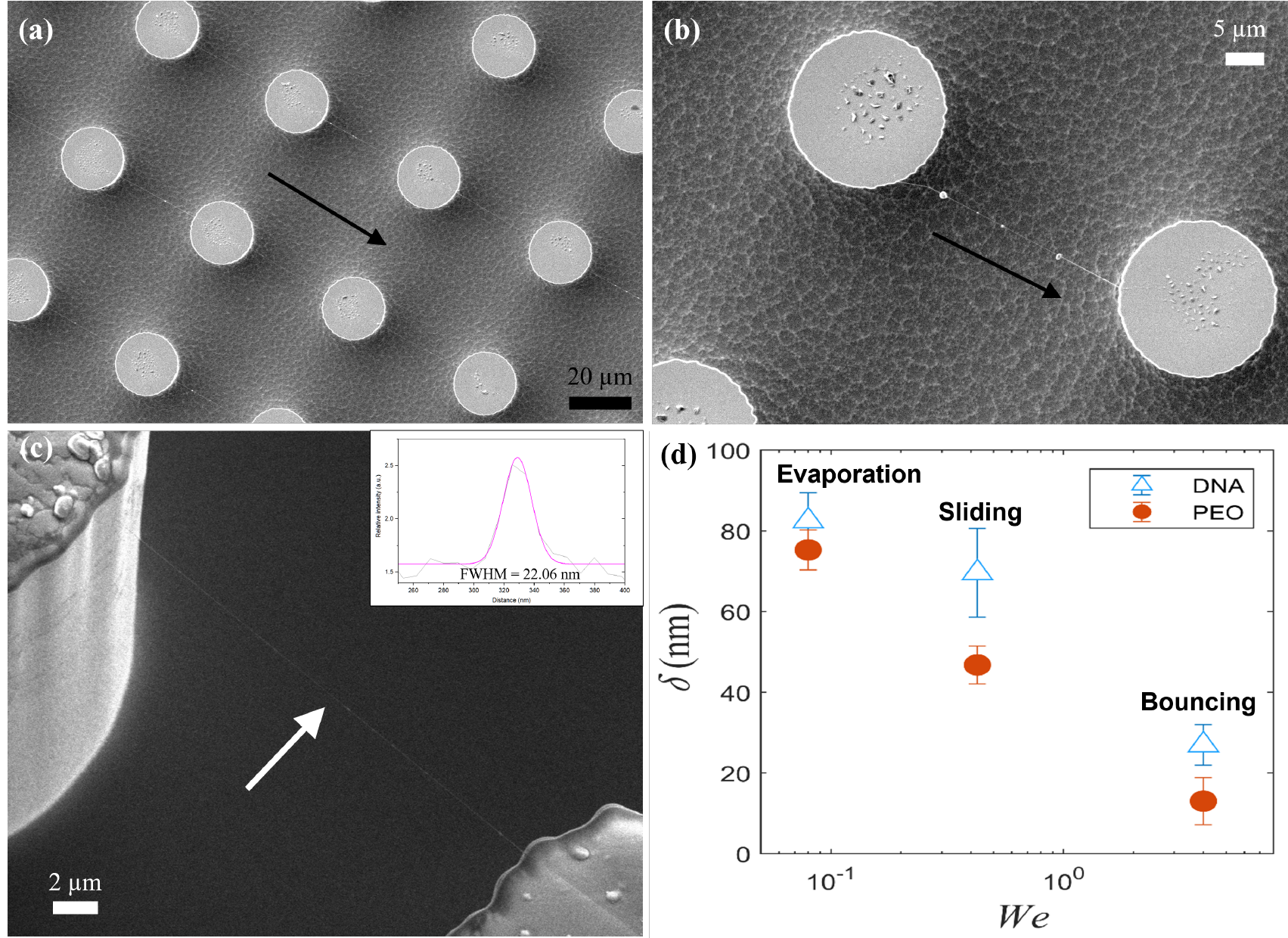}
	\caption{{\bf SEM images of dry polymer fibrils} deposited on the top of micro-pillars after the bouncing impact of the drops on a $45^o$ inclined substrate. 
(a) Suspended and stretched PEO fibrils stretched between all pillars.  The fibrils are plasma-coated with 2 nm iridium layer.
(b) Zoom-in of a PEO fibril attached between two pillars.
The arrows in (a) \& (b) point in the rebounding direction of the drops.
(c) High-magnification SEM image of DNA filament (white arrow), with a best fit of the intensity profile in the inset.
(d) Variations of PEO and DNA filament thickness with different drop impact condition.  The lowest $We$ is estimated for drop evaporation and the middle data for sliding drops, while the highest $We$ corresponds to the impact results.
	Each data point is the average of three measurements from different filaments. 
	Error bars denote the range of the measurements.}
	\label{Fig_005}
\end{figure*}
In contrast to the almost fixed-spaced filaments from impacts on the micro-pillars, the filaments generated by impacts on the randomly-spaced roughness-peaks of the Glaco-coated surfaces do not stay separate, as many of them are generated quite close to each other.   
The high-speed videos show that adjacent filaments often merge to form a unique branching structure, which was shown in Fig. \ref{Fig_1}(a).  
When filaments are too close to each other on the free surface, their conical menisci are pulled together by the surface tension to minimize surface energy, akin to the {\it ``Cheerios effect''} \cite{Vella2005}.  
%\textcolor{red}{Vella, D.; Mahadevan, L. (2005). "The Cheerios effect". American Journal of Physics. 73 (9): 817–825}
This is shown in the sequence of frames in Fig. \ref{Fig_03}(b) and Supplemental video 4.  
Once the cones merge, between 2nd and 3rd frame, two filaments meet at a cuspy corner.  
This is an unstable configuration, as is well-known in the evolution of foam, i.e., the stable angle between the filaments (Plateau borders) in a static 2-D configuration must be $120^{\circ}$.  %In Supplementary Fig. 1, 
The initial angle between the two bottom branches are smaller than this value and the nodal point is rapidly pulled downwards to increase this angle.  
The downward zipping speed of the nodal point is as high as \textcolor{black}{3.29} m/s, as marked by the arrow in Fig. \ref{Fig_03}(b).
This merger can occur multiple times in sequence to form a distinctive branching structure.  The last frame in the sequence points at five levels of nodes leading to the same filament attaching to the drop surface. % as shown in the video sequence corresponding to Fig. \ref{Fig_03}(b).  

%This action of the post-merge node, which was measured to be $U_{merge}$ = 0.39 m/s.
%which for reference is of similar order as the impact velocity $U$ = 0.37 m/s, even though they are not dynamically connected.

% Further experiments with plasma-treated polymer surfaces (Methods) indicate that air-film-mediated bouncing also persists for (super)hydrophobic surfaces exhibiting nano-roughness for roughness amplitudes below 100 nm (Supplementary Movie 12). This observation suggests that the benefits of air cushions can be exploited in combination with nano-texturing of liquid-repellent surfaces—for example, to minimize undesired solid–liquid contact even further.

%%%%%%%%%%%%%%%%%%%%%%%%%%%%%%%%%%%%%%%%%%%%%%%%%%%%%%%%

Our goal is to deposit polymer filaments between adjacent pillars to allow their study using \textcolor{black}{molecular characterizations}. 
Therefore after being pulled out they must attach to the next pillar and not be pulled out vertically as in Fig. \ref{Fig_1}(b).  
Figure \ref{Fig_04}(b) shows how this is accomplished for impacts on an inclined substrate.
The filament (marked by a red arrow) is pulled from pillar marked 1, but as the free surface is pulled towards pillar 2
it attaches to this pillar and is left behind, free-standing between these two pillars, while a new filament is pulled from pillar 2 and so forth.  
This attachment between adjacent pillars occurs during the retraction phase of the pancake, 
but not during the final rebound of the drop where it moves primarily up and away from the pillars.  Regular deposition of these filaments between pillars is therefore limited to a sub-region of the area of the maximum impact footprint.  This process is clearly seen in the Supplementary video clip 5.

Soon after these impacts the substrates are removed and subjected to SEM imaging to characterize the size and structure of the deposited filaments. 
Figure \ref{Fig_005} shows typical SEM images of filaments attached between the tops of adjacent pillars.
In panel (b) three tiny nodules indicate remnants of micro-droplets on the filament before drying.
In (a) regular filaments are aligned towards bottom right between all pillars.
This coincides with the inclination angle of the substrate, i.e. filaments are pulled out in the bouncing direction. \textcolor{black}{Supplementary Fig. S2} shows even larger number of regularly deposited filaments.
In the inset in panel (c) we fit the intensity profile across one filament to determine the Full-Width-Half-Max diameter, giving here a thickness value of $\delta = 22$ nm.  This thickness measurement was performed on numerous filaments and the results are shown in panel \ref{Fig_005}(d), where  
we contrast our drop-impact method to earlier work using drop-evaporation or rolling drops \cite{de2011, marini2014raman, gentile2012, marini2015}.  
This also includes results from drops containing $\lambda -$DNA.
This shows that our impact method can generate thinner filaments than the other two approaches, which we verified with evaporating and rolling drops under the same experimental conditions as the impacts.
The filaments span approximately 30 $\mu$m distance between the pillars, which is much longer that the 16 $\mu$m DNA segments. From Fig. \ref{Fig_005}(d), we therefore conclude that filaments are formed from bundles of overlapping unpaired $\lambda-$filaments.
Due to the wide range of possible bouncing regimes, the attached location and thickness of the DNA filaments on the pillars could be optimized further by controlling the droplet impact condition and micro-pillar shapes.

\textcolor{black}{Using Raman spectroscopy (see Supplemental Material) we confirmed that the B-form DNA is the favoured conformation in nano-filaments at $45$\% humidty. 
During the drop bouncing, the DNA molecules are stretched by the hydrodynamic forces, which where described in the previous report. 
Below $80$\% humidity the stable structure is the B form, as verified in several experiments \cite{gentile2012, marini2015}.} 
% \section*{Discussion and Conclusions}

In summary, our experimental results present a simple but robust method to deposit stretched polymer fibers between micro-pillars, using inpacting drops. This deposition occurs much more rapidly than for the drying technique.  While drying uses the entire drop liquid, leaving a heap of polymers at the center, the bouncing drop can be collected after the rebound, to bounce again on a new substrate, thereby using much smaller sample volume of liquid.  Furthermore, the stronger stretching during the dynamic impact, appears to produce thinner filaments than the rolling technique, which can be beneficial for their detailed electron micro-scope imaging.

\section*{Methods}
{\it Imaging and drops:}
Supplementary Fig. S1 shows a schematic of the experimental setup.
We use Phantom V2511 ultra-fast CMOS video cameras at up to 100,000 frame-per-second (fps), with a Leica Z16 APO long-distance microscope with adjustable magnification of up to 29.2, which gives pixel resolution up to about 1 $\mu$m/px.
In some experiments we used two synchronized cameras, for different viewing angles.

The drops are released by gravity from a glass nozzle with 100 $\mu$m inner diameter.
Liquid was fed using a syringe pump at slow flow-rate of 2 $\mu$l/min to generate repeatable drops through static pinch-off.

{\it Micro-pillar fabrication:}
\textcolor{black}{Silicon wafers (4-inch diameter and 500 $\mu$m thickness) with a 2.4 $\mu$m thickness silica layer were used as substrates. The arrays of cylindrical pillars (diameter: 20 $\mu$m, height: 50 $\mu$m, and pitch: 50 $\mu$m) were fabricated by photolithography and dry etching. After fabrication, the wafers with features were cleaned by Piranha solution (${H_{2}SO_{4}} : {H_{2}O_{2}}= 3 : 1$ by volume) for 15 min at 110 $C^{\circ}$, and then flushed by de-ionized water and dried in a Spin Rinse Dryer. 
To achieve hydrophobicity, these wafers were then coated with perfluorodecyltrichlorosilane (FDTS) by molecular vapor deposition (ASMT 100E).}

{\it PEO:}
The PolyEthylene-Oxide (PEO) was purchased from Sigma-Aldrich ($M_{w}$ = 4 × 10$^{6}$g/mol).  It was in powder form, which was mixed with DI water (Milli-Q Plus system) using a magnetic mixer for at least 10 hours with low angular speeds before use.  
% in Fig. \ref{Fig_05}(a). 

{\it SEM:}  
The PEO polymer filaments stretched between the pillars are plasma-coated with 2 nm iridium to increase the contrast and prevent it from breaking from the SEM electric beam, which was usually kept below 3 kV. 5 kV was used in Fig. 5 (c) and a magnification of 3500. Twice the thickness of the iridium layer was subtracted from the diameter measurements.

{\it $\lambda-$DNA:}
was purchased from Thermo Fisher Scientific with molecular weight of $31.5 \times 10^{6}$ DA and concentration of 300 ppm from Thermo Fisher Scientific. 
The solvent is mostly water, but contains 10 mM Tris-HCl (pH 7.6) and 1 mM EDTA.
%It has typical molecular length of 30 $\mu$m ????
The 48.5-kb lambda DNA molecule has a B-form contour length of $\sim$16.5 $\mu$m (with an upper limit of 0.34 nm/bp) \cite{wang1998}. 

{\it Raman Spectroscopy:}
Laser-confocal Raman microscopy (WiTec Apyron) was used to deploy the Raman measurements.
To avoid breaking the PEO nano-filaments, a low dose (5 mW) 532-nm laser was used as the excitation source. A $100\times$ objective lens (NA = 0.9) was used to focus the laser spot on DNA nano-filaments and collect Raman scattering signal.

\section*{Acknowledgments}
This work was supported by King Abdullah University of Science and Technology (KAUST),
under grants URF/1/2126-01-01 \& BAS/1/1352-01-01. \\
\textbf{Data availability}\\
%\subsection*{Data availability}
Source data are available for this study. 
% The data used to generate Fig.~\ref{mural1}d \textcolor{blue}{and Fig.~\ref{plot}, and Supplementary Fig.~2 can be found in the Supplementary Data 1 and 2 respectively.}
All other data that support the plots within this paper and other findings of this study are available from the corresponding author upon reasonable request.\\
\textbf{Competing interests}\\
%\subsection*{Competing interests}
The authors declare no competing interests.\\
\textbf{Author contributions}\\
S.T.T. and Z.Q.Y. conceived the whole project, designed the research \& participated in all phases of the work.
Z.Q.Y. took the initial high-speed video data; 
S.T.T. and E.D.F. provided the original idea for DNA;
Z. Q. Y. performed most experiments, participated in PEO, DNA preparation, high speed imaging and data analysis. 
A. A. J. carried out some of the experiments; 
H. M., M. S. and Z.P. designed and fabricated the micro pillar substrates; 
Z.Q.Y. and Z.P. performed the SEM imaging;
Z.P. worked on Raman spectroscopy measurements and analysis;
Z.Q.Y. and S.T.T. wrote the manuscript and all authors discussed the results and edited the paper.  
% STT acknowledges early discussions with T. G. Etoh and K. Takehara.  This work was supported by King Abdullah University of Science and Technology (KAUST).
% For your review copy (i.e., the file you initially send in for
% evaluation), you can use the {figure} environment and the
% \includegraphics command to stream your figures into the text, placing
% all figures at the end.  For the final, revised manuscript for
% acceptance and production, however, PostScript or other graphics
% should not be streamed into your compliled file.  Instead, set
% captions as simple paragraphs (with a \noindent tag), setting them
% off from the rest of the text with a \clearpage as shown  below, and
% submit figures as separate files according to the Art Department's
% instructions.

% \clearpage

% \noindent {\bf Fig. 1.} Please do not use figure environments to set
% up your figures in the final (post-peer-review) draft, do not include graphics in your
% source code, and do not cite figures in the text using \LaTeX\
% \verb+\ref+ commands.  Instead, simply refer to the figure numbers in
% the text per {\it Science\/} style, and include the list of captions at
% the end of the document, coded as ordinary paragraphs as shown in the
% \texttt{scifile.tex} template file.  Your actual figure files should
% be submitted separately.
\bibliography{scibib}
\bibliographystyle{Science}

\end{document}